\documentclass[11pt]{article}
 \usepackage{amsfonts}
 \usepackage{amssymb}
 \def\bsh{\backslash}
 
 \def\adt{\dot \alpha}

 \newfont{\bbbold}{msbm10 scaled \magstep1}

 \def\bbA{\mbox{\bbbold A}}
 
 \def\bbC{\mbox{\bbbold C}}

 \def\bbR{\mbox{\bbbold R}}

 \def\cA{{\cal A}}

 \def\cN{{\cal N}}
 \def\cO{{\cal O}}

 \newfont{\goth}{eufm10 scaled \magstep1}

 \def\gs{\mbox{\goth s}}
 
 \def\gu{\mbox{\goth u}}

 \def\a{\alpha}

 \def\d{\delta}
 \def\e{\epsilon}
 \def\vf{\varphi}

 \def\l{\lambda}\def\L{\Lambda}
 \def\m{\mu}

 \def\th{\theta}
 \def\X{\Xi}
 \def\be{\begin{equation}}\def\ee{\end{equation}}
 \def\bea{\begin{eqnarray}}\def\eea{\end{eqnarray}}
 \def\ba{\begin{array}}\def\ea{\end{array}}
 \def\x{\xi}
 \def\o{\omega}\def\O{\Omega}
 \def\del{\partial}

 \def\xz{\times}

 \def\del{\partial}

 \let\la=\label

 {}

 \def\nn{\nonumber}
 \def\bd{\begin{document}}
 \def\ed{\end{document}}
 \def\bea{\begin{eqnarray}}\def\barr{\begin{array}}\def\earr{\end{array}}
 \def\eea{\end{eqnarray}}
 \def\ft#1#2{{\textstyle{{\scriptstyle #1}\over {\scriptstyle #2}}}}
 \def\fft#1#2{{#1 \over #2}}
 \newcommand{\eq}[1]{(\ref{#1})}
 \def\eqs#1#2{(\ref{#1}-\ref{#2})}
 \def\det{{\rm det\,}}
 \def\tr{{\rm tr}}\def\Tr{{\rm Tr}}

\begin{document}

 \thispagestyle{empty}

 \hfill{KCL-TH-00-48}

 \hfill{hep-th/0008047}

 \hfill{\today}

 \vspace{20pt}

 \begin{center}
 {\Large{\bf Chiral Superfields in IIB Supergravity}}
 \vspace{30pt}

 {P. Heslop and P.S. Howe}
\vskip 1cm {Department of Mathematics} \vskip 1cm {King's College,
London} \vspace{15pt}

 \vspace{60pt}

 {\bf Abstract}

 \end{center}

The field strength superfield of IIB supergravity on $AdS_5\xz S^5$
is expanded in harmonics on $S^5$ with coefficients which are $D=5,
N=8$ chiral superfields. On the boundary of $AdS_5$ these
superfields map to $D=4,N=4$ chiral superfields and both sets of
superfields obey additional fourth-order constraints. The
constraints on the $D=4,N=4$ chiral fields are solved using
harmonic superspace in terms of prepotential superfields which
couple in a natural way to composite operator multiplets of the
boundary $N=4,D=4$ superconformal field theory.

{\vfill\leftline{}\vfill \vskip  10pt

 \baselineskip=15pt \pagebreak \setcounter{page}{1}


A key feature of the AdS/CFT correspondence \cite{mal} is the simple fact that
the symmetry group of $AdS$ spacetime is the same as the conformal
group of the boundary Minkowski space, a feature which extends to
supersymmetric theories in a natural way when they are formulated
in superspace. Perhaps the clearest and best-established example of
the correspondence is between IIB supergravity on $AdS_5\xz S^5$
and $D=4,N=4$ supersymmetric $SU(N_c)$ Yang-Mills theory (in the
large $N_c$ limit) on the boundary of $AdS_5$. The Kaluza-Klein
spectrum of this supergravity theory was found some time ago \cite{gun},
and in \cite{hw} a natural family of gauge-invariant composite operators
in N=4 SYM was written down using a formulation of the theory in a
certain harmonic superspace \cite{hh2}. It was later shown in \cite{af} that these
two sets of supermultiplets are in one-one correspondence. In this
note we shall look at the relation between these two sets of
supermultiplets in a very explicit fashion by tracing what happens
to the linearised IIB field strength supermultiplet on $AdS_5\xz
S^5$ as one first expands it in terms of $S^5$ harmonics and then
passes to the boundary to obtain a set of $D=4, N=4$ chiral
superfields. We show that the constraints satisfied by this set of
field strength superfields can easily be solved in harmonic
superspace in terms of a family of Grassmann-analytic prepotential
superfields (one for each chiral field) and the picture is
completed by coupling the prepotentials to generalised current
multiplets. The constraints on the latter, arising from the gauge
invariances of the prepotentials, are precisely those that are
obeyed by the family of $N=4$ SYM composite operators mentioned
above.

The basic geometrical set-up is summarised by the following
diagram:

\be
\ba{ccc} AdS^{5,5|32} & \rightsquigarrow & M^{4|16}\xz S^5 \\ &&\\
\Big\downarrow && \Big\downarrow \\ &&\\ AdS^{5|32}
&\rightsquigarrow & M^{4|16} \ea \ee

where the squiggly arrows denote passing to the boundary. Each of
these superspaces is a coset space of the supergroup $PSU(2,2|4)$,
with the notation indicating the (even$|$odd) dimensions. Thus
$AdS^{5,5|32}$ denotes the superspace whose body is $AdS_5\xz S^5$
and which has 32 odd dimensions while $AdS^{5|32}$ is the $D=5,
N=8$ superspace which has $AdS_5$ for its body and which has  32
odd dimensions. The boundary space on the bottom row is $D=4, N=4$
Minkowski superspace. The isotropy group of $AdS^{5,5|32}$ is
$Spin(1,4)\xz USp(4)$ while that of $AdS^{5|32}$ is $Spin(1,4)\xz
SU(4)$ and the former fibres over the latter with fibre $S^5$. The
boundary superspaces are most easily described by viewing
$PSU(2,2|4)$ as the $D=4, N=4$ superconformal group. The generators
of the corresponding Lie superalgebra are $\{D,P,K,M,N|Q,S\}$
standing for dilations ($D$), translations ($P$), Lorentz
transformations ($M$), internal $SU(4)$ transformations ($N$), and
$Q$ and $S$-supersymmetry transformations, all in four dimensional spacetime. The istropy group of
$M^{4|16}$ is generated by $\{D,K,M,N|S\}$ while the isotropy group
of $M^{4|16}\xz S^5$ is generated by $\{D,K,M,N'|S\}$, where $N'$
denotes the generators of the $USp(4)$ subgroup of $SU(4)$.


We recall that IIB supergravity has the following bosonic component
fields: the graviton, a complex scalar, a complex two-form
potential, and a real four-form potential whose five-form
field-strength is self-dual. The fermions are a complex Weyl
gravitino and a complex Weyl spinor. In the linearised theory these
fields can be put together in a chiral superfield $\bbA$ satisfying
a fourth-order constraint which is schematically of the form
$D^4\bbA\sim \bar{D^4 }\bar{\bbA}$ \cite{hw2}. In fact, the $D^4$ component of
$\bbA$ can be decomposed into two fields which are irreducible
tensors of $SO(1,9)$ with 770 and 1050 components. The former is
the Weyl tensor and is real while the latter is purely imaginary.\footnote{This fact was observed by N. Berkovits (private communication); it can also be derived from the full theory.}
The five-form field strength only enters into the chiral superfield
at $\th^4$ level via its derivative and the resulting six-index
tensor, antisymmetric and self-dual on five of the indices, but not
antisymmetric on all six, transforms under the 1050 representation
of $SO(1,9)$. In the full theory, there is still a complex scalar
superfield but it takes its values in the coset space $U(1)\bsh
SL(2,\bbR)$. The geometrical field strength tensors, the torsion
and curvature and three-form and five-form field strengths,
constructed from the supervielbein, the connection and the
superspace two- and four-form potentials, are described in terms of
a spinor superfield $\L$ whose leading component is the spinor
field of the supergravity multiplet and a five-index antisymmetric
tensor superfield $G_{abcde}$ whose leading component is the
covariantised spacetime five-form field strength. These superfields
are of course not independent since together they describe the
supergravity multiplet.

The homogeneous superspace $AdS^{5,5|32}$ has been described in references \cite{mt,krr,ctrrz,orrt}.
Since it is a coset one can use standard homogeneous space
techniques to construct the components of the torsion and curvature
tensors. In fact, it turns out that for this superspace the spinor
superfield $\L$ vanishes while the superfield $G_{abcde}$ has
constant components which are non-vanishing only if all of the
indices are in the range 0 to 4 or in the range 5 to 9, the
non-vanishing components being proportional to five-dimensional
epsilon tensors. If one expands about this background, then to
first order, the theory is still described by a chiral scalar
superfield. Moreover, since the anticommutator of two fermionic
covariant derivatives has the same form as in flat space (with the
difference that the spacetime derivative must be amended), it
follows that $D^4 \bbA$ still falls into the same two
representations of the Lorentz group as before, and it is not
difficult to show that the same fourth-order constraint holds.

As we remarked above $AdS^{5,5|32}$ can be regard as a fibre bundle
over $AdS^{5|32}$ with fibre $S^5$. This means that we can expand
out the IIB chiral superfield $\bbA$ in harmonics on the
five-sphere. The expansion has the form

\be
\bbA=\sum_{k=0}^{\infty} n^{R_1}\ldots n^{R_k} \cA^{(k)}_{R_1\ldots
R_k} \ee

Here $n^R, R=1,\dots 6$, is a unit vector on the sphere and each of
the tensors $\cA^{(k)}$ is symmetric and traceless on its $SO(6)$
indices.

The component superfields $\cA^{(k)}$ are all chiral superfields on
$AdS^{5|32}$ and satisfy fourth-order reality conditions. However,
for the components with indices the spinorial covariant derivative
is altered by a term involving the $SO(6)$ connection of the fibred
superspace. Clearly this term depends on the representation the
component is in. The lowest component superfield $\cA^{(0)}$ is the
field strength superfield of linearised $N=8,D=5$ gauged
supergravity.

We can pass to the boundary in the following way (see \cite{krr,orrt}
for related work). For $AdS^{5|32}$ we may choose the coset representative $s(r,\vf,x,\th)$ to be of the form

\be
s=e^{\vf\cdot S} e^{\ln r D} g(x,\th). \label{section}
\ee

Here $g(x,\th)$ is the standard coset representative for $D=4, N=4$
Minkowski superspace, $\vf$ denotes the odd coordinates associated
with $S$-supersymmetry, and $(r,x)$ are coordinates for $AdS_5$. In
these coordinates the metric of $AdS_5$ itself is

\be
ds^2=r^{-2}\left( dr^2 + dx^2\right) 
\ee

where $dx^2$ denotes the flat $D=4$ Minkowski space metric. We will
take the boundary to be at $\vf=0$,  $r\rightarrow 0$. 

$AdS^{5|32}$ has isotropy group $Spin(1,4)\xz SU(4)$, with Lie
algebra generated by $\{\frac 12 (K-P), M, N\}$ The remaining coset
generators are given by $\{\frac12(K+P),D|Q,S\}$.    
When $\vf=0$
\be
ds s^{-1}
 =\frac 1{r}e^a P_a + \frac1{\sqrt{r}}(e^{\a_q }
Q_{\a_q}) + (d\vf^{\a_s}S_{\a_s})+\frac{dr}{r}D
\label{viel}
\ee  
where $e^A=(e^a,e^{\a_q})$ is the standard one form basis in flat
$\cN=4$ superspace, given by:
\be 
        dgg^{-1}=e^aP_a + (e^{\a_q}Q_{\a_q}).
\ee
From (\ref{viel}) we see that the one form basis of $AdS^{5|32}$ when
$\vf = 0$ is
\be
E^a=\frac 1{r}e^a \qquad
E^4= \frac{dr}{r}\qquad
E^{\a_q}=\frac1{\sqrt{r}}e^{\a_q}\qquad
E^{\a_s}=d\vf^{\a_s}.
\ee

We wish to compare the bulk superisometries of $AdS^{5|32}$ with
superconformal transformations of $\cN=4$ Minkowski space on the
boundary. Here we introduce the following indices: $M=(m,\m_q)$,
$M'=(m,4,\m_q,\m_s)$, and similarly for the flat indices $A$,
$A'$. Superconformal
transformations act on the coordinates $z=(x,\th)$ by $\d_{sc} z^M =
\x^M$, where
\be
g {\bf T} g^{-1} = \x^a P_a  + (\x^{\a_q} Q_{\a_q}) + (\o^{\a_s}S_{\a_s})
+ \l D + ... \label{conf}
\ee
Here ${\bf T}$ is an arbitrary infinitesimal element of the Lie
algebra, and $\x^M=\x^A e_A{}^M$. The dots represent other elements of
the isotropy Lie algebra. 
The isometries of $AdS^{5|32}$ act on the coordinates
$Z'=(x,r,\th,\vf)$ of $AdS^{5|32}$ by $\d_{ads} Z^{M'}
=\X^{M'} = \X^{A'} E_{A'}{}^{M'}$ where 
\be
s {\bf T} s^{-1} = \X^{A'}{\bf T}_{A'} + \O^r{\bf T}_r.
\ee
On the right hand side we have split the Lie
algebra up into a coset part and an isotropy group part for
$AdS^{5|32}$. Using (\ref{section}) and (\ref{conf}) we find that at
$\vf = 0$
\be 
s{\bf T} s^{-1}
                        = \frac 1{r} \x^a P_a +
                        \frac1{\sqrt{r}} (\x^{\a_q}Q_{\a_q}) +
                        \sqrt{r} ({\o}^{\a_s}S_{\a_s}) + \l D +
                        ... 
\ee 

This gives us the following at $\vf=0$
 \be
\ba{rcl}
        \d_{ads} x        &=& \d_{sc} x + O(r^2) \\
        \d_{ads} \th      &=& \d_{sc} \th\\
        \d_{ads} \vf      &=& \sqrt{r} \, \o \\
        \d_{ads} \ \ r &=&r \,\l .
\ea
\ee

The $O(r^2)$ term appears from changing
from the superconformal basis to the AdS basis of the Lie algebra.
We see that the action of the isometries of the bulk space on the
coordinates $(x,\th)$ is equivalent to 
superconformal transformations on $N=4$ super Minkowski space. Also the
action of the isometries transverse to this boundary disappears, and
so setting $\vf=0$ is a consistent thing to do. (Note that this is
not the case if we instead choose the coset representative $s=
e^{\ln r D} e^{\vf\cdot S}g(x,\th)$. In this case one has to argue
that the
dependence on $\vf$ of the bulk fields tends to zero.)

A similar construction holds for $AdS^{5,5|32}$ except that we also
have to include coordinates for $S^5$. In the boundary limit $r
\rightarrow 0$, $\vf=0$ the
dependence of the chiral superfields on half of
the odd coordinates disappears and we thus end up with a family of
$D=4, N=4$ chiral superfields $\{A^{(k)}_{R_1\dots R_k}\},
k=0,1,2,\ldots$. These superfields obey the chirality constraint

\be
\bar D_{\adt}^i A^{(k)}=0 \ee

where we have adopted two-component spinor notation for
four-dimensional spacetime and where the index $i$ runs from 1 to
4. The fourth-order constraints become

\be
 D^4_{\{R_1 R_2} A^{(k)}_{R_3\ldots R_{k+2}\}}=
 \overline{D^4_{\{R_1 R_2} A^{(k)}_{R_3\ldots R_{k+2}\}}}
 \la{d4con}
\ee

where the curly brackets denote traceless symmetrisation. The
derivative $D^4_{RS}$ is formed from $(D_{\a i})^4$ by taking the
component which is a Lorentz scalar and which transforms under the
real 20-dimensional representation of $SU(4)$; in $SO(6)$ notation
this projection corresponds to a second-rank symmetric traceless
tensor. The lowest superfield in this set, $A^{(0)}$, is the field
strength superfield of $D=4, N=4$ conformal supergravity. Note that
while the $N=8, D=5$ superfields $\cA^{(k)}$ are on-shell fields
the $N=4, D=4$ superfields $A^{(k)}$ are off-shell.

In order to complete the picture we now need to translate the
fourth-order constraint \eq{d4con} into harmonic superspace
language \cite{gikos1}. The harmonic superspace we shall use is $M\xz S(U(2)\xz
U(2))\bsh SU(4)$ where $M$ denotes $N=4$ super Minkowski space. We
shall employ the standard GIKOS trick of working on $M\xz SU(4)$
and demanding that the behaviour of the fields with respect to the
isotropy group is fixed. We write $u\in SU(4)$ as
$u_I{}^i=(u_r{}^i,u_{r'}{}^i)$ in index notation, and denote the
inverse of $u$ by $u_i{}^I=(u_i{}^r,u_i{}^{r'})$. The capital $I$
index is understood to be acted on by the isotropy group and thus
splits naturally into $r=1,2$ and $r'=3,4$. With the aid of $u$ and
its inverse we can convert $SU(4)\ i$ indices into isotropy group
$I$ indices and vice versa. Differentiation on the group itself is
done using the right-invariant derivatives $D_I{}^J$, with
$D_I{}^I=0$. From the fact that these derivatives generate the left
action of the group on itself we find

\be
D_I{}^J u_K{}^k= \d_K{}^J u_I{}^k -\frac14 \d_I{}^J u_K{}^k \ee

The set of right-invariant derivatives splits into the isotropy
group derivatives $\{D_r{}^s, D_{r'}{}^{s'},D_o\}$ and coset
derivatives $\{D_r{}^{s'},D_{r'}{}^s\}$. The isotropy derivatives
have been written in $\gs\gu(2)\oplus\gs\gu(2)\oplus\gu(1)$
notation (so that $D_r{}^r=D_{r'}{}^{r'}=0$), and our normalisation
is such that

\be
D_o u_r{}^i=\frac12 u_r{}^i\qquad D_o u_{r'}{}^i=-\frac12
u_{r'}{}^i \ee

The coset space $S(U(2)\xz U(2))\bsh SU(4)$ is a compact complex
manifold with complex dimension four (it is the Grassmannian of
two-planes in $\bbC^4$), and the derivatives $D_r{}^{s'}$ can be
thought of as the components of the $\bar\del$ operator on this
space while the derivatives $D_{r'}{}^s$ are the complex
conjugates. The introduction of harmonics allows the possibility of
having superfields which satisfy generalised chirality constraints;
in the present case, the derivatives $D_{\a r}=u_r{}^i D_{\a i}$
and $\bar D_{\adt}^{r'}=\bar D_{\adt}{}^i u_i{}^{r'}$ anticommute
with each other and commute with $D_r{}^{s'}$. A field $F$ on
harmonic superspace  which satisfies

\be
D_{\a r}F=\bar D_{\adt}^{r'}F=0 \ee

is called Grassmann analytic, or G-analytic, while a field which
satisfies $D_r{}^{s'}F=0$ is called harmonic analytic, or
H-analytic. A field which is both G- and H-analytic will be called
an analytic superfield.

For the following discussion it will be useful to define the object
$u^{ij}=-u^{ji}$ by

\be
u^{ij}:=\frac12 \e^{rs} u_r{}^i u_s{}^j \ee

The fact that $u\in SU(4)$ then implies that

\bea
 u_{ij}&:=&\frac12 \e_{ijkl} u^{kl} \nn\\
        &=&\frac12 \e_{r's'}u_i{}^{r'}u_j{}^{s'}
\eea

We shall also employ the following abbreviations for fourth-order
Grassmann derivatives which are Lorentz scalars,

\bea
 D^4&:=&(D_{\a r})^4\qquad D'^4:=(D_{\a r'})^4\nn \\
 \bar D^4&:=&(\bar D_{\adt}{}^{r})^4\qquad
 \bar D'^4:=(\bar D_{\adt}^{r'})^4
\eea

The final technicality we shall need is the notion of harmonic
conjugation \cite{gikos1} which combines a generalisation of the antipodal map
for $\bbC P^1$ to the coset $S(U(2)\xz U(2))\bsh SU(4)$ with
complex conjugation. We shall denote this operation by a tilde. For
the $u$'s we have

\bea
 \tilde{(u_r{}^i)}&=& u_i{}^{r'}\nn \\
 \tilde{(u_{r'}{}^i)}&=& -u_i{}^r
\eea

while for any $u$-independent object we use ordinary complex
conjugation. It follows that G-analyticity and H-analyticity are
both preserved by harmonic conjugation and that

\bea
 \tilde{(u^{ij})}=u_{ij}
\eea

Returning to the boundary chiral field strength superfields, we can
define the following family of objects

\bea
  C^{(k)}:&=&u^{i_1 j_1}\ldots u^{i_{k+2}j_{k+2}} D^4_{i_1 j_1,
i_2 j_2} A^{(k)}_{i_3j_3,\ldots i_{k+2} j_{k+2}}\nn \\
          &\sim&  u^{i_1 j_1}\ldots u^{i_{k}j_{k}}
          D^4 A^{(k)}_{i_1j_1,\ldots i_{k} j_{k}}
\eea

where we have converted the $SO(6)$ indices $R$ into antisymmetric
self-dual pairs of $SU(4)$ indices and then contracted all these
pairs with $u^{ij}$'s. We claim that each $C^{(k)}$ is analytic.
H-analyticity is immediate, since $u^{ij}$ is annihilated by
$D_r{}^{s'}$. G-analyticity follows because the presence of the
$u$'s implies that the projection of $D^4$ involves only the
derivatives $D_{\a r}$. Since the product of five such derivatives
vanishes identically, we see that $D_{\a r}C^{(k)}=0$, while $\bar
D_{\adt}^{r'}$ can be anti-commuted past the $D$'s to act on the
chiral field $A$ whereupon it gives zero. The fourth-order
constraint on the field $A^{(k)}$ (with $k$ $SO(6)$ indices) can
therefore be written in the form

\be
C^{(k)}=\tilde C^{(k)}. \ee

Since this constraint is non-dynamical it can be imposed by a
Lagrange multiplier superfield, $V^{(k)}$, say, which may be taken
to be G-analytic. The free action for $A^{(k)}$ can therefore be
written\footnote{Similar methods can be used to find prepotentials in Minkowski superspace, as in \cite{hst}}

\be
S^{(k)}=\left(\int d^4x\,d^8\th\,\frac12 (A^{(k)})^2 \right) + c.c.
+ \int d\m\, V^{(k)}(C^{(k)}-\tilde C^{(k)}) \ee

In this formula, the first integral is an $N=4$ chiral integral
while the second one is a harmonic superspace integral. The
harmonic superspace measure $d\m$ is defined by

\be
d\m:=d^4x\,du\, D'^4 \bar D^4 \ee

where $du$ denotes the standard measure on the coset space. If we
vary the above action with respect to $A^{(k)}$, which is here
regarded as an otherwise unconstrained chiral superfield, we get an
expression for $A^{(k)}$ in terms of the prepotential $V^{(k)}$. It
is

\be
 A^{(k)}_{i_1j_1,\ldots i_kj_k}=\int du\, u_{i_1 j_1}\ldots
 u_{i_k j_k}\bar D^4 V^{(k)}
\ee

This formula generalises in a natural way the expression for the
chiral $N=2$ Maxwell field strength multiplet in terms of a
G-analytic charge two potential \cite{gios}.

We note that $u^{ij}$ has $U(1)$ charge 1 so that $C^{(k)}$ has
$U(1)$ charge $k+2$. Since the charges of $D^4,D'^4,\bar D^4,\bar
D'^4$ are $2,-2,-2$ and $2$ respectively the charge of the harmonic
measure is $-4$, so that $V^{(k)}$ has charge $-k+2$. We note
further that, since $C^{(k)}$ is both G- and H-analytic, the action
$S^{(k)}$ has an abelian gauge invariance of the form

\be
 V^{(k)}\rightarrow V^{(k)}+ D_r{}^{s'} X^{(k)}_{s'}{}^r
\ee

where the parameter superfield $X^{(k)}$ is G-analytic.

The prepotentials $V^{(k)}$ couple naturally to generalised
currents $\cO_{k+2}$ via the Noether coupling

\be
 S^{(k)}_{{\rm Noeth}}=\int d\,\m V^{(k)} \cO_{k+2}
\ee

Clearly the current $\cO_{k+2}$ has the same properties as the
corresponding constraint $C^{(k)}$, in particular $\cO_k$ has
$U(1)$ charge $k$.

The currents can be realised explicitly as composite operators in
$N=4$ $SU(N_c)$ Yang-Mills theory. We recall that the $N=4$ Maxwell
field strength supermultiplet is described in $N=4$ super Minkowski
space by the Sohnius superfield $W_{ij}$ \cite{sohn} which is antisymmetric and
self-dual and which satisfies

\bea D_{\a i} W_{jk}&=&D_{[\a i} W_{jk]}\nn\\ \bar{D}_{\adt}^i
W_{jk}&=& -\frac23 \d^i_{[j} \bar D_{\adt}^l W_{k]l} \eea

These constraints can be translated into harmonic superspace by
introducing the field $W:=u^{ij} W_{ij}$ \cite{hh2}. This is an analytic
superfield which has $U(1)$ charge 1 ($D_o W=W$) and which is real
with respect to harmonic conjugation, $W=\tilde W$. Conversely one
can show that that such an analytic superfield is equivalent to the
Sohnius superfield. In the non-Abelian theory the field strength
superfield $W$ is H-analytic, but the G-analyticity condition has
to be made covariant with respect to the gauge group. However, the
gauge-covariant operators $\cO_k:=\tr (W^{k})$ are analytic in the
usual sense, and so provide the desired family of generalised
currents which couple to the prepotentials $V^{(k)}$.

{\bf Acknowledgement}

This research was supported in part by PPARC SPG grant 613.

 \end{document}